# Carrier and Phonon Dynamics in Multilayer WSe$_2$ Captured by Extreme Ultraviolet Transient Absorption Spectroscopy

*Juwon Oh,*[1,2,†] *Hung-Tzu Chang,*[1,3,†] *Christopher T. Chen,*[4] *Shaul Aloni,*[4] *Adam Schwartzberg,*[4] *Stephen R. Leone*[1,5,6]*

[1]Department of Chemistry, University of California, Berkeley, California 94720, USA
[2]Department of Chemistry, Soonchunhyang University, Asan, 31538, Korea
[3]Max Planck Institute for Multidisciplinary Sciences, 37077 Göttingen, Germany.
[4]Molecular Foundry, Lawrence Berkeley National Laboratory, Berkeley, California 94720, USA
[5]Chemical Sciences Division, Lawrence Berkeley National Laboratory, Berkeley, California 94720, USA
[6]Department of Physics, University of California, Berkeley, California 94720, USA

**ABSTRACT**

Carrier and phonon dynamics in a multilayer WSe$_2$ film are captured by extreme ultraviolet (XUV) transient absorption (TA) spectroscopy at the W N$_{6,7}$, W O$_{2,3}$, and Se M$_{4,5}$ edges (30-60 eV). After the broadband optical pump pulse, the XUV probe directly reports on occupations of optically excited holes and phonon-induced band renormalizations. By comparing with density functional theory calculations, XUV transient absorption due to holes are identified below the W O$_3$ edge whereas signals at the Se $M_{4,5}$ edges are dominated by phonon dynamics. Therein, 0.4 ps hole relaxation time, 1.5 ps carrier recombination time, and 1.7 ps phonon heating time are extracted. The acquisition of hole and phonon-induced signals in a single experiment can facilitate the investigation of the correlations between electron and phonon dynamics. Furthermore, the simultaneous observation of signals from different elements can be further extended to explore photochemical processes in multilayers and alloys, thereby providing key information for their applications in electronics, photocatalysts, and spintronics.

## Introduction

Semiconducting transition metal dichalcogenides (TMDs) have attracted much scientific interest in recent years as an active material of photocatalytic systems, photovoltaics, optoelectronics and photodetectors based on their unique optical, spin and electronic properties.[1-4] TMDs, generally expressed as $MX_2$ (M = transition metal, such as Mo and W, and X = chalcogen, such as S, Se and Te), are two-dimensional nanomaterials composed of covalently bonded X-M-X sheets that are stacked together through van der Waals interactions. Each layer may harbor defects, which can serve as a tool for modifying electronic properties, optical responses, and catalytic activities.[5-7] Moreover, due to the layered structure, they show important layer-dependent properties, e.g. the crossover from an indirect band gap in multiple-layered TMDs to a direct band gap in a monolayer with spin-valley coupling.[1]

Based on these promising features, monolayer TMDs have been comprehensively investigated for their photoluminescent nature, strong excitonic effects, and high possibility for spin control.[4] Recently, this attention has been extended to multilayer TMDs with their several advantages.

Because of the low energy threshold for atomic defects of each TMD layer, the stability of monolayer TMDs is sensitive to the exposed environmental conditions.[8,9] Although the interlayer interaction leads to an indirect bandgap, the preserved excitonic absorption bands of multiple layered TMDs allow for strong light absorption, which is favorable to applications in photovoltaics, photocatalysts, and photodetectors.[2,10] Furthermore, recent advancements in the preparation of TMDs allows for mass production in the form of multilayers,[11,12] which ensures that multilayer TMDs are being newly illuminated as practically applicable nanosheet materials.

With this interest, the photophysics and carrier dynamics of multilayer TMDs have been extensively investigated with optical time-resolved spectroscopies.[13-20] However, observed spectra in the visible (VIS), near-infrared (NIR), IR and terahertz region generally exhibit spectral congestion and the signals are often without carrier specificity. Particularly, because most bulk TMDs are non-photoluminescent, the spectral congestion in these optical spectroscopic analyses hinders deconvolution of carrier and phonon features and extraction of carrier dynamics. Here, core-level spectroscopy can provide a breakthrough for investigating carrier dynamics in multilayer TMDs. Electronic transitions from atomic core orbitals by X-ray or extreme ultraviolet (XUV) light bring element specificity to spectral analyses.[21-23] In addition, core-level transient absorption (TA) spectroscopy, consisting of optical pump and X-ray or XUV probe, can monitor carrier distributions by disentangling electron and hole signals as a function of energy.[18,24-29]

With the advantages of core-level spectroscopy, here we explore multilayer $WSe_2$ with XUV TA spectroscopy. Among TMDs, $WSe_2$ has also attracted attention due to its ambipolar nature and high conductivity and as an effective layered heterojunction material for optoelectronics, spintronics, and moiré excitons.[30,31] Despite its significance, there are relatively few studies for $WSe_2$ compared to $MoS_2$, $MoSe_2$, and $WS_2$. Also, most studies have been focused on monolayer and heterolayer systems.[32] Herein, we study the photoinduced carrier and phonon dynamics and their impact on XUV spectral signatures in multilayer $WSe_2$. Our broadband XUV TA measurement allows observing the core-level transitions at the W $N_{6,7}$, W $O_{2,3}$, and Se $M_{4,5}$ edges (30-60 eV). Density functional theory calculations are utilized to identify the measured spectral signatures. Furthermore, a comparative analysis with $WS_2$ and the observation of both W and Se edges facilitate disentangling hole relaxation, carrier recombination and phonon heating in $WSe_2$, where the significant gap in the carrier relaxation time between $WS_2$ and $WSe_2$ reflects different carrier-phonon coupling strengths despite their high similarity in electronic and crystal structures.

## Experimental and Computational Methods

**Sample synthesis.** The XUV transient absorption experiments were carried out on 40 nm thick $WSe_2$ films deposited on 30 nm thick $Si_3N_4$ windows. Using plasma-enhanced chemical vapor deposition (PECVD), $Si_3N_4$ was deposited with a thickness of 16 nm to passivate the Si edge frame of the $Si_3N_4$ window, which prevents the formation of silicon selenide in the process of selenation.[33] Second, a $WO_3$ film was synthesized with plasma-enhanced atomic layer deposition on the $Si_3N_4$ window.[34] The thickness of the $WO_3$ film was calculated according to the required thickness of $WSe_2$ using the ratio of the density between the two assuming no W loss in the subsequent selenation. The substitution of oxygen with selenium was conducted in a 600 °C tube furnace for one hour using $H_2Se$ (5 sccm) flowed with Ar (100 sccm) as buffer gas. The sample was then characterized with X-ray photoelectron spectroscopy (XPS) and Raman spectroscopy (see Supporting Information).

**XUV transient absorption experiment.** To investigate the photoexcited carrier and phonon dynamics in WSe$_2$, the sample was excited (pumped) by a 4 fs long optical pulse with a spectrum spanning 500-1000 nm. The subsequent XUV probe pulse was produced by high harmonic generation of a sub-4 fs optical pulse in a Kr or Ar gas jet. The choice of gas medium for high harmonic generation was determined by the required XUV spectral range. Kr was used for 30-50 eV (W N$_{6,7}$ and O$_{2,3}$ edges) and Ar for 50-70 eV (Se M$_{4,5}$ edge) photon energies. The experiments were conducted with 4-8 mJ/cm² VIS-NIR fluences, corresponding to peak intensities of 0.5-1.0 TW/cm². The resulting excited carrier density ranges from 0.9 x 10$^{20}$ - 1.7 x 10$^{20}$ cm$^{-3}$. Details of the experimental setup can be found in Ref. 19 and 25. The transient absorption at time delay $t$ is defined as $\Delta A(t) = A_{pumped}(t) - A_{pumped}(t = -50\,fs)$. Here, $t$ indicates the arrival time difference between the XUV and the optical pulses. Because the lifetime of W 4f$_{7/2,5/2}$ core excitons are much shorter than 50 fs,[19] $\Delta A(t \leq -50\,fs)$ is expected to be zero.

**Computational methods.** To understand the XUV transient absorption results, core-level absorption spectra of WSe$_2$ were simulated with density functional theory (DFT) using the full-potential linearized augmented plane wave (FP-LAPW) method implemented in the Elk code.[35] Ground state DFT calculations including spin-orbit coupling were carried out with 12x12x3 k-points under the local spin density approximation.[36] The core-level absorption spectra were subsequently computed with a random phase approximation.

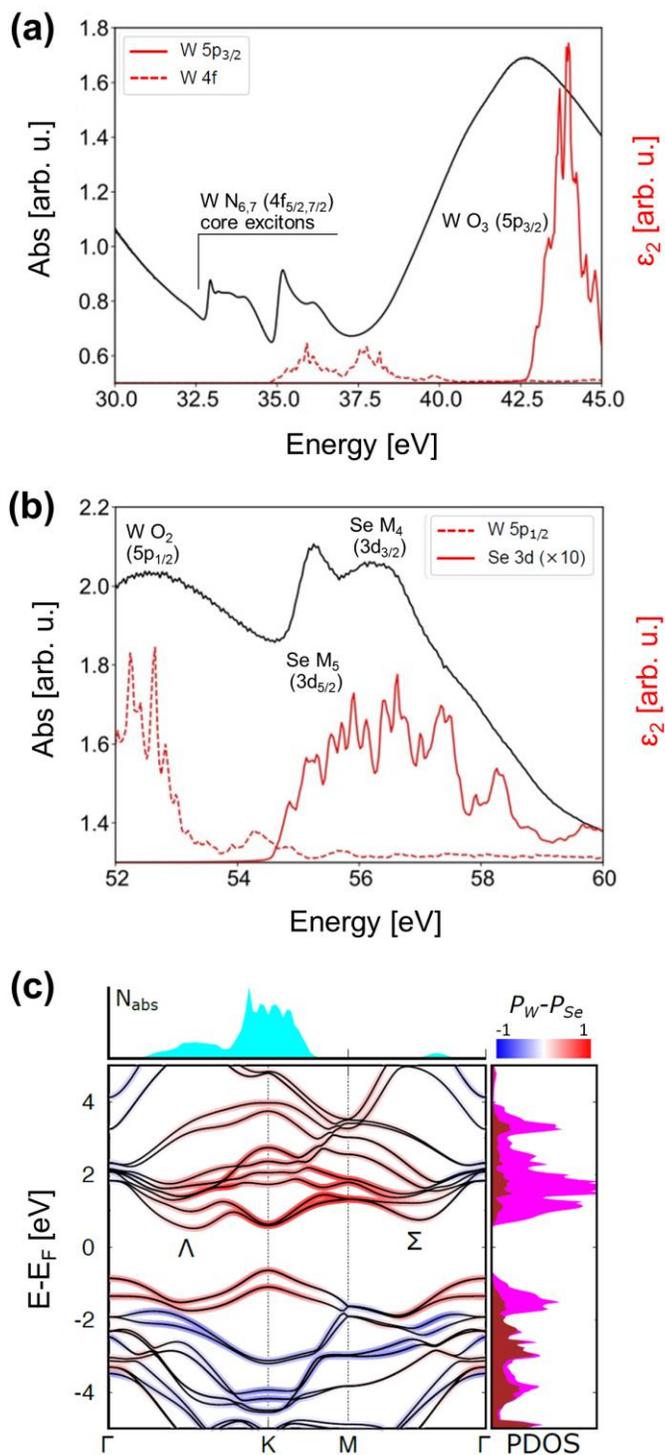

**Figure 1**. Measured static absorption spectra (black lines) of WSe$_2$ at the W N$_{6,7}$ and O$_3$ edges (a) and at the Se M$_{4,5}$ edge (b). Simulated absorption due to transitions from the W 4f, 5p, and Se 3d

levels are shown in red. Here $\varepsilon_2$ denotes imaginary part of the dielectric function. (c) displays the band structure of WSe$_2$, the orbital character of the bands, the projected density of states of W 5d (magenta) and Se 4p (brown) orbitals, and the estimated number of carriers due to absorption of the optical pulse ($N_{abs}$). The method used to estimate the initial distribution of photoexcited carriers is detailed in Ref. 19.

## Results and Discussion

**Static XUV absorption spectra.** The core-level absorption (*Abs*) from W and Se edges are shown in Figures 1a and 1b, respectively. The absorption of W N$_{6,7}$ and O$_{2,3}$ edges exhibit two absorption peaks along with fine structure detail between 32 and 37 eV and an intense absorption edge starting at approximately 38 eV (Figure 1a). The absorption from W core levels is highly similar to the spectrum of WS$_2$,[19] and the absorption peaks within 32-37 eV are thus identified as core-excitons from the W 4f core orbitals and the absorption above 38 eV from the W 5p$_{3/2}$ levels. The Se M$_{4,5}$ edge is situated at 55-58 eV (Figure 1b) and the broad absorption peak between 51-54 eV is due to absorption from the W O$_2$ edge.

To verify the assignments, we compare the measured spectrum with DFT calculations (Figures 1a and 1b, red lines). A global shift of +6 eV is applied to the computed spectra to compensate the underestimated gap between the core levels and the conduction band in DFT. The measured Se M$_{4,5}$ absorption qualitatively agrees with the simulated spectrum. Note that the magnitude of the Se edge absorption is smaller than the neighboring W O$_2$ edge. This is due to the lack of Se 4p orbital character close to the conduction band minimum (Figure 1c). An energy mismatch of approximately 2 eV between the simulation and the measurement is observed at the W N$_{6,7}$ and O$_3$ edges. This can be explained by the excitonic effect in the core-level excitations. The conduction band minima are mainly contributed by W 5d orbitals and the oscillator strengths for core excitations to the conduction band minima are much larger for W 4f and 5p levels than Se 3d orbitals. As excitons are typically dominated by transitions at the band edges, the lowering of transition energies due to Coulombic interactions between the excited electron and the core hole is more likely to occur at the W 4f and 5p transitions than at the Se $M_{4,5}$ edges.

**Carrier dynamics at the W $N_{6,7}$ and $O_3$ edges.** Optically excited electrons and holes can directly modify core-level absorption through creation or elimination of excitation channels, which would result, in principle, in positive (core-to-valence band (VB)) or negative (core-to-conduction band (CB)) signals in XUV TA spectra, respectively.[18,19,26-29] This "state-blocking" effect is governed by the oscillator strengths between the core levels and the occupancy of the VB and CB, where the photoexcited carriers reside. As the valence and conduction band edges are dominated by W 5d orbitals (Figure 1c), XUV TA due to state blocking is expected to occur at transitions from W core levels.

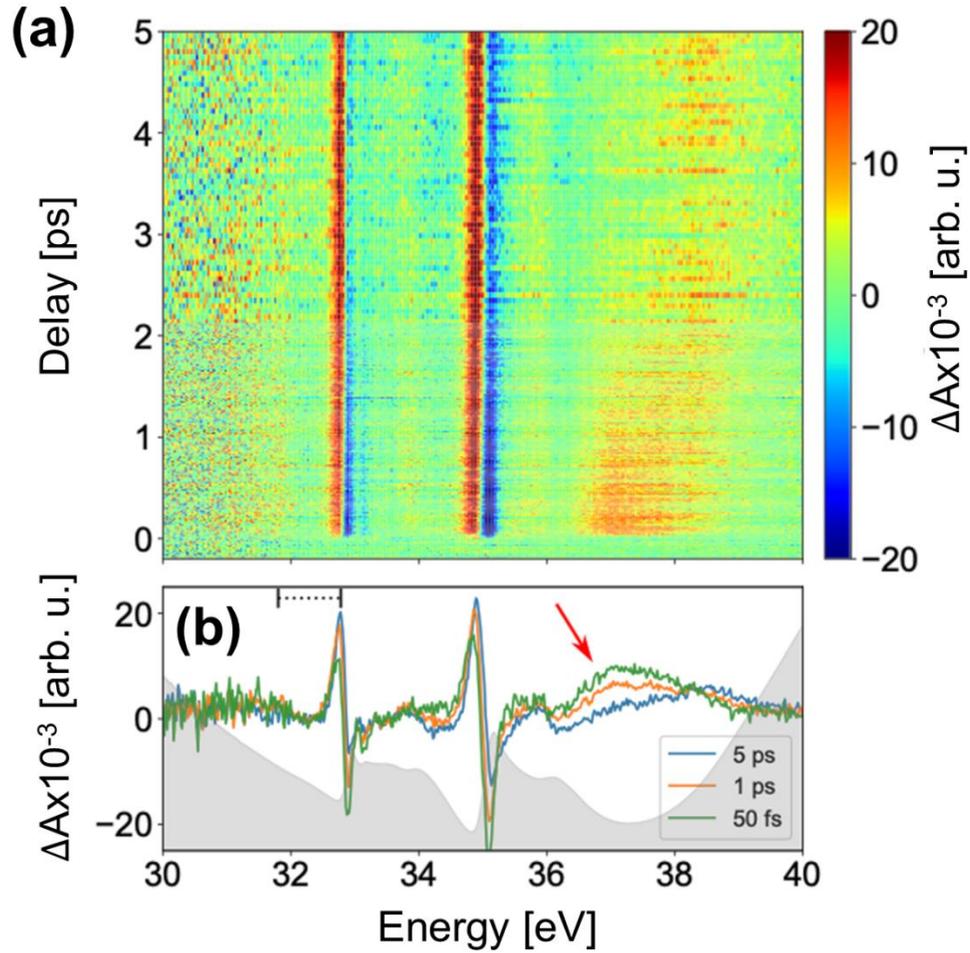

**Figure 2.** (a) XUV transient absorption at the W $N_{6,7}$ and $O_3$ edges between -200 fs and 5 ps time delays. (b) Lineouts of transient absorption spectra at 50 fs, 1 ps, and 5 ps delays. The static spectrum is shown in gray as a reference.

Figure 2 displays the XUV TA spectra at the W $N_{6,7}$ edge (33-36.5 eV) and $O_3$ edge (37-40 eV). We first focus on TA spectrum at 50 fs time delay (green line in Figure 2b). As the carrier relaxation due to electron-phonon coupling typically occurs from 100 fs timescale,[37] TA signals at 50 fs delay are mainly due to photoexcited carriers. Strong derivative features occur at W 4f core-exciton transitions (~33 eV and ~35 eV) in the TA spectrum, which are due to photoinduced changes in both valence shell occupations and screening.[19] A positive change in absorbance is observed below the W $O_3$ edge (~37 eV, red arrow in Figure 2b), which we tentatively assign to the holes in the valence band.[19] Another weak positive feature that is barely above noise level occurs ~1.5 eV below the onset of the W $N_7$ edge (31.5 eV). This feature may also be due to holes in the valence band but is too weak to be analyzed. To verify our assignments, we consult the DFT-simulated transient absorption spectrum.

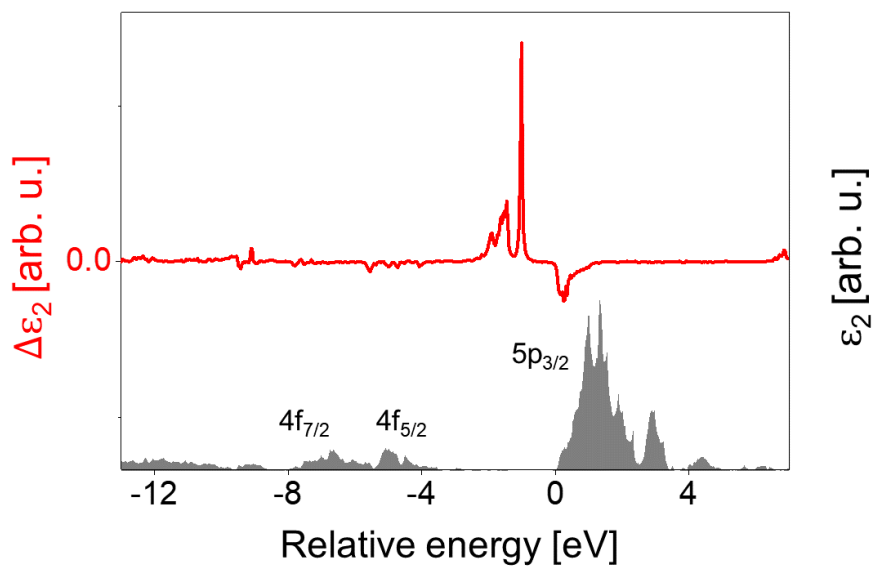

**Figure 3**: Simulated transient absorption spectrum at the W $N_{6,7}$ edges and the W $O_3$ edge. The transient absorption is the difference between the core-level absorption spectrum with a thermalized carrier distribution and the ground state spectrum. Details of the simulation are described in the Supporting Information.

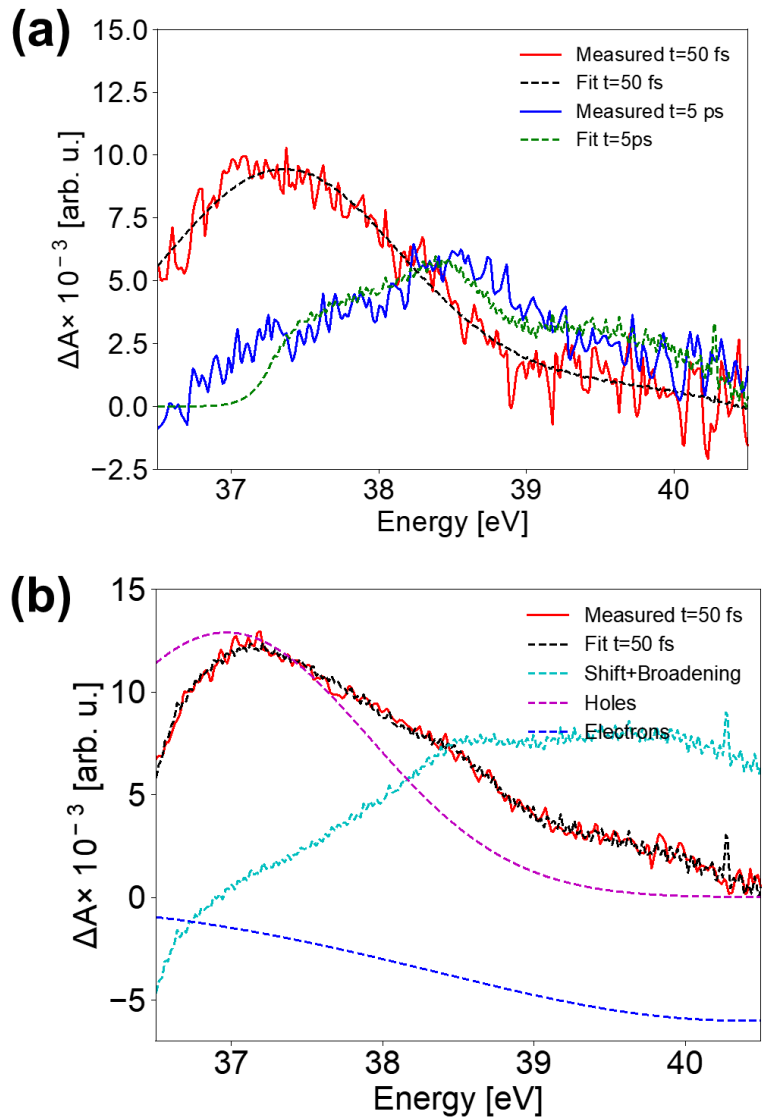

**Figure 4**: (a) Transient absorption spectra in the W $O_3$ edge region at 50 fs and 5 ps delays and fitting profiles with spectral shifting and broadening of the static spectrum plus carrier state-blocking. (b) Decomposition of W $O_3$ edge transient absorption at 50 fs delay with shifting and broadening of the static spectrum and electron and hole state blocking features represented by two Gaussian functions. Assuming the oscillator strengths of core-level transitions are uniform throughout the W $O_3$ edge, the integrated area of the positive and negative state-blocking features over the full energy range are set to equal in the fitting.

The computed TA spectrum (red line in Figure 3) exhibits a strong positive feature 1-2 eV below the W $O_3$ edge due to holes in the VB. The electrons in the CB contribute to the negative feature above the onset of the W $O_3$ edge. The occurrence of the positive feature 1-2 eV below the edge onset is in good agreement with the experimental results. In addition, the simulated TA within the W $N_{6,7}$ edge is barely discernible and much weaker than at the W $O_3$ edge. This indicates that the oscillator strengths between the W 4f levels and the VB and CB edge are much weaker than the transitions from W 5p levels. The lack of oscillator strength between the W 4f levels and the VB maximum explains the lack of positive features due to optically excited holes below the W $N_{6,7}$ edge. Finally, the negative feature due to electrons at the W $O_3$ edge is weaker than the corresponding positive feature due to holes. This combined with the red shift of the absorption edge induced by carrier-induced band gap lowering,[18,19] which would lead to an increase in absorbance, can explain the absence of a negative change of absorbance in the experimental results (Figure 2, 39-40 eV).

The positive feature around 38 eV exhibits two different dynamical behaviors (Figure 2). In the 37-38 eV region, the feature decays with increasing time delay. By contrast, the absorption increases with time delay in the neighboring 38-39 eV region. While it is tantalizing to interpret this as the relaxation of holes towards the band edge, the calculated photoexcitation of carriers in the band diagram (Figure 1c) indicates that the majority of the photoexcited holes resides in the K valley close to the VB edge and the pump pulse is unlikely to create holes 2 eV below the valence band edge. To identify and assign the spectral components, we thus conduct a global fitting of the TA signals in the W $O_3$ edge region (36.5-40.5 eV) with singular value decomposition (SVD).[19] The SVD results show that the dynamics at 36.5-40.5 eV contain two major spectral components (Figure S2). The component with the largest singular value (Figure S2d) exhibits a broad positive signal with a maximum at 38.2 eV, which corresponds to the positive maximum of TA spectrum at 5 ps. The spectrum of this component lacks the derivative feature typical for carrier state-blocking. Instead, such positive feature occurring directly at the band edge is characteristic of the phonon heating effect.[24] The second largest spectral component (Figure S2e) shows the positive and negative features at 37 and 38.8 eV, which can be interpreted as features of holes and electrons, respectively.

Although the SVD analysis enables the qualitative assignment of two dynamical components, it is ill-disposed for quantitative analysis of spectral changes containing energy-shifting, narrowing, and broadening, which may occur in the TA results due to carriers relaxing to the band edge and core-level energy shifts due to changes in electronic screening and phonon dynamics, for example.[24,38,39] Therefore, we conduct an additional fitting of the TA spectra at the W $O_3$ edge with spectral shift, broadening, and carrier-induced state blocking to further distinguish between the different contributions.[24] Figure 4a shows the fitting results of TA spectra at 50 fs and 5 ps delays, which is in good agreement with experimental results. The TA spectrum at 5 ps can be described by the red-shifted static spectrum with -0.01 eV plus a broadening of 0.2 eV without any carrier state-blocking. This is well matched with the SVD results and indicates band renormalization by the heating effect. The discrepancy below 37.2 eV is due to the artificial removal of the static absorption from the tail of W $N_6$ edge (Figure S3). The TA spectrum at 50 fs is reproduced with the shifting and broadening of the static spectrum and electron and hole state-blocking signals (Figure 4b). Here the negative feature assigned to electrons in the conduction band is much broader than the hole feature. In addition, the central energy of the negative feature is ~3 eV above the positive hole feature. This phenomenon can be potentially explained by the broadening of the W $5p_{3/2}$ transitions (Figure 1a). The broadening of the core-level absorption edge may not only broaden the transient absorption signal, but also lead to transient absorption features appearing at higher energies into the extended edge regions. For example, when many-body interactions between the core-level-to-CB and VB-to-CB excitations become non-negligible, the static extended edge absorption may include contributions from both core-level-to-CB and VB-to-CB transitions.[39,40] In this case, the optically excited electrons occupying the orbitals above the Fermi level not only affect the one-particle transitions from the core level but also the many-body transitions consisting of both core-level transitions and valence excitations, leading to TA signals in the extended edge region. Note that such an effect is not applicable to the hole feature as energy conservation dictates that in the pre-edge region, only the mixing of core-to-VB and inner-VB transitions can have major contributions when many-body effects are significant.

The two different TA spectral analyses allow us to assign the positive signal between 36.5-38 eV as due to holes. The dynamics of photoexcited holes can thus be extracted from the positive feature in the TA spectra at ~37.5 eV (red arrow in Figure 2b). First, the central energy of the feature, $E_{mean}$, shifts from low- to high-energy with increasing time delay. $E_{mean}$ as a function of time delay can be fitted by a single exponential with a time constant of $0.4 \pm 0.2$ ps (Figure 5a). The energy shift of the hole feature reflects the relaxation of the hole population towards the VB maximum. Because the photoexcitation mainly occurs in the K valley (Figure 1c), the relaxation may involve intervalley redistribution of holes into the Γ valley, which has similar energy to the K valley. The hole state-blocking signal diminishes at long time limit (Figure 5b) and the dynamics can be fitted with a single exponential decay of $1.5 \pm 0.1$ ps. As the orbital character at the VB edge is dominated by the W 5d orbitals, where transitions from the W 5p core levels are allowed, the disappearance of the hole state-blocking signal is interpreted as the recombination of holes with electrons in the CB.

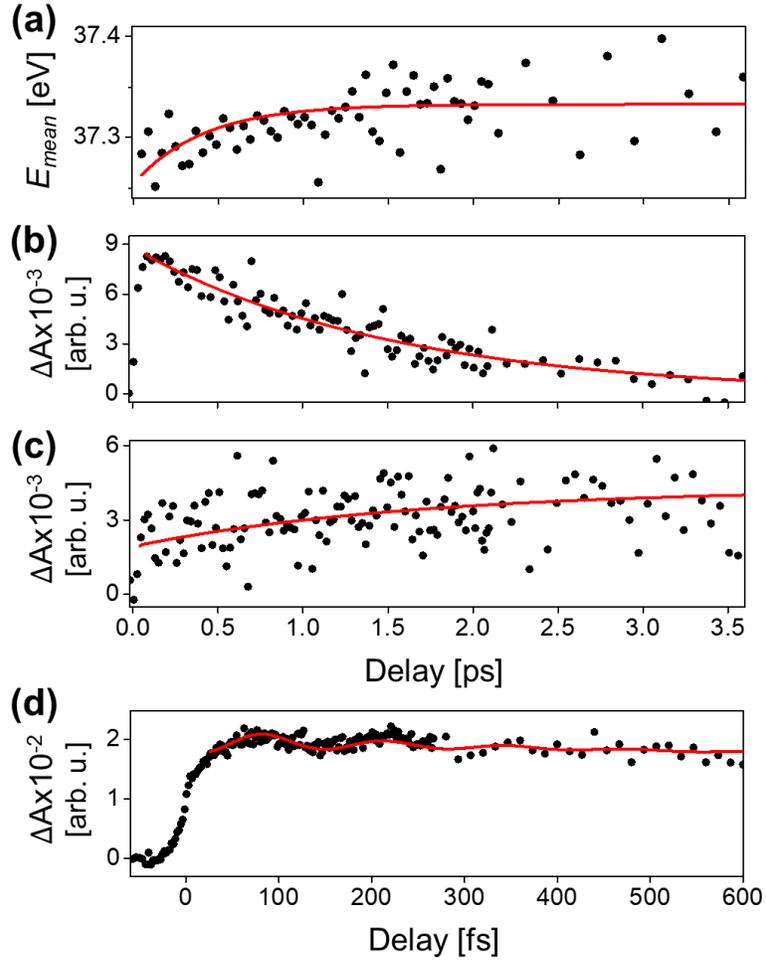

**Figure 5**: (a) Temporal lineouts of spectral mean energy, $E_{mean} = (\int E\Delta A\, dE)/(\int \Delta A\, dE)$, of the TA signals between 36.7 and 37.8 eV with a time constant of $0.4 \pm 0.2$ ps. Temporal lineouts of the averaged TA signals in the range of (b) 36.7-37.8 eV and (c) 38.3-39.5 eV exhibit time constants of $1.5 \pm 0.1$ and $1.7 \pm 0.4$ ps, respectively. (d) Temporal lineouts of the averaged TA signals in the range of 34.6-34.9 eV with 131 fs-period damped oscillation curve (red line).

**Phonon dynamics at W $N_{6,7}$ and $O_3$ edge.** The spectral analyses indicate that the positive TA signal at 38 eV can be solely described by shifting and broadening of the static spectrum. As the signal persists at >3 ps delays where the carrier state-blocking signal already decays to zero, indicating a completion of carrier recombination, we assign this feature as due to phonon-induced band renormalization. The amplitude of the XUV TA spectra within the 38.3-39.5 eV range exhibits a single-exponential rise (Figures 2b and 5c), where the fitted time constant is 1.7 ± 0.4 ps. The TA signals at the W $N_{6,7}$ edges also show a spectral blue shift at a similar timescale (Figure S4). Here, the fitted time constant of 1.7 ps matches with the timescale of Bragg peak suppression measured with ultrafast electron diffraction.[41] This agreement suggests that the TA signals at the $N_{6,7}$ and $O_3$ edges are caused by phonon heating of $WSe_2$. In addition, the similarity between the carrier recombination time and the phonon heating time implies that non-radiative carrier recombination also contributes to the lattice heating process.

In addition to the incoherent phonon heating dynamics, weak oscillatory signals (Figures 5d and S5) are observed at ~34.7 eV atop the core-exciton TA signal due to carriers.[19] The observed oscillations have a ~131 fs period, which is consistent with the intense Raman signal at 254.4 cm$^{-1}$ (Figure S1c) and agrees with the ~7.5 THz coherent $A_{1g}$ optical mode measured with optical pump-probe spectroscopy.[20] Nevertheless, the involvement of the $E_{2g}^1$ mode cannot be ruled out due to its proximity in mode frequency.[42] The <1 ps lifetime of the oscillation here is much shorter than the >1 ps coherent phonon decay time observed with optical techniques.[20] This may be explained by the low amplitude of the oscillatory signal, which is close to noise level. Moreover, as the core-excitons are more strongly influenced by the carrier-induced change of valence shell occupation and screening at short time scale and phonon-induced band renormalization at long time scale, the dynamics of these two components and their crossover may also influence the amplitude of the coherent-phonon-induced oscillations, which share the same spectral region. Therefore, the dephasing time of the coherent phonons cannot be inferred from the temporal lineout measured here.

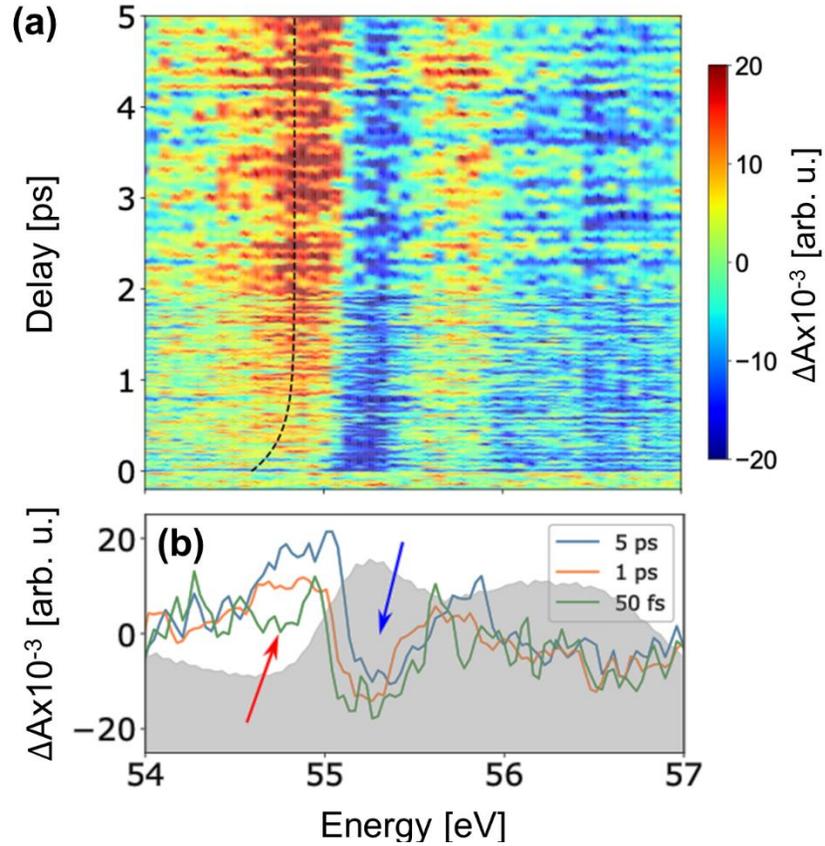

**Figure 6**. (a) XUV transient absorption at the Se $M_{4,5}$ edges between -200 fs and 5 ps time delay. The black dashed line denotes the fitted spectral mean energy, $E_{mean} = (\int E \Delta A \, dE)/(\int \Delta A \, dE)$, of the positive feature (red arrow in (b)) between 54.5-55.1 eV. (b) displays the lineouts of transient absorption spectra at 50 fs, 1 ps, and 5 ps delays. The static spectrum is shown in gray as a reference.

**Shifts in the Se M$_{4,5}$ edge.** The XUV TA spectra at the Se M$_{4,5}$ edge presents several long-lived features up to 5 ps delay (Figure 6). Here we focus on the features within 54.5-55.5 eV at the Se M$_5$ edge, which do not overlap with transient absorption signals from the Se M$_4$ edge transitions. Between 54.5-55.1 eV, a positive feature increases in magnitude with increasing time delay (red arrow in Figure 6) and a negative change of absorbance is present between 55.1-55.5 eV that slowly decays with time (blue arrow in Figure 6).

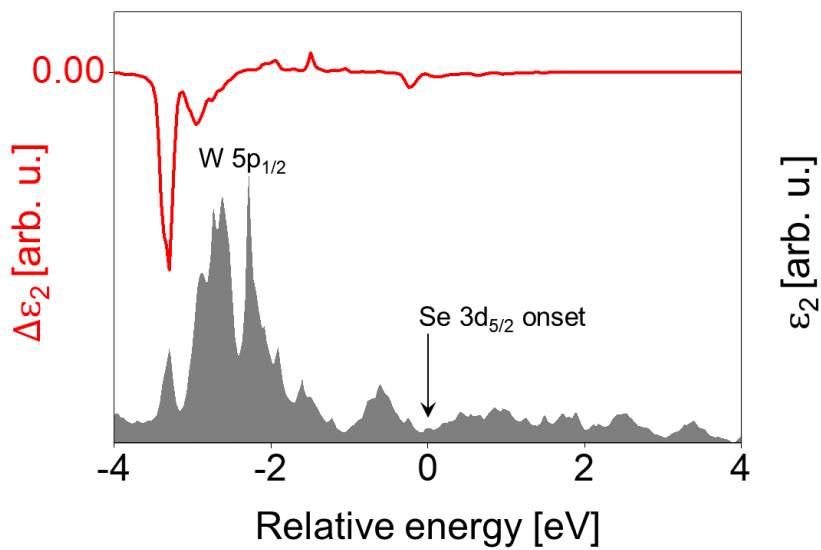

**Figure 7.** Simulated transient absorption spectrum at Se M$_{4,5}$ edge. Details of the simulation are described in Supporting Information.

The positive and negative features centered at ~54.8 eV and ~55.3 eV, respectively, may be assigned to photoexcited holes and electrons in the valence shell. However, as the onset of the Se M$_5$ edge is at 55 eV, the hole signal is expected at ≤54 eV, rather than 54.8 eV, according to the bandgap of bulk WSe$_2$.[43] In addition, the amplitude of the positive signal is close to zero at 50 fs time delay and slowly increases with time. This contradicts the expectation for holes, whose population should decay with time due to recombination. The projected density of states of WSe$_2$ shows that the valence band edge is predominantly contributed by the W 5d orbitals, whereas the VB with significant Se orbital character is mainly >1 eV below the VB edge (Figure 1c). Therefore, if the positive signal is directly due to holes, it should decay with time both because of recombination and relaxation to the valence band edge, where the holes become invisible to the Se 3d core excitations, rather than increase in magnitude. Furthermore, the simulated core-level TA spectrum within the independent particle approximation (Figure 7) shows that the magnitude of the TA signal at the Se M$_{4,5}$ edge is very small compared to the neighboring W 5p transitions. This is consistent with the simulations carried out by Schumacher et al. on MoSe$_2$, a similar TMD material.[44] In their work, Schumacher et al. also showed that many-body effects are negligible for Se M$_{4,5}$ edge transitions, thereby validating the qualitative comparison of the computation under independent particle approximation with the experimental results. The temporal behavior of the signals at the Se M$_{4,5}$ edge and the comparison with DFT simulations indicate that these transient absorption features are not directly due to state blocking of carriers.

To understand the spectral features, we consult their dynamics. The center of the positive feature between 54.5-55.1 eV shifts from low to high energy with time (black dashed line in Figure 6). This shift can be fitted with a single exponential with a time constant of $0.4 \pm 0.2$ ps. This timescale coincides with the 0.4 ps carrier relaxation observed at the W O$_3$ edge. The rise of the positive feature at 54.5-55.1 eV can be described by a single exponential with a time constant of $1.6 \pm 0.3$ ps (Figure 8a). This is similar to both the recombination time extracted at the W O$_3$ edge and the phonon population rise time observed previously with ultrafast electron diffraction.[41] As the feature is not due to carrier state blocking, we attribute it to phonon dynamics.

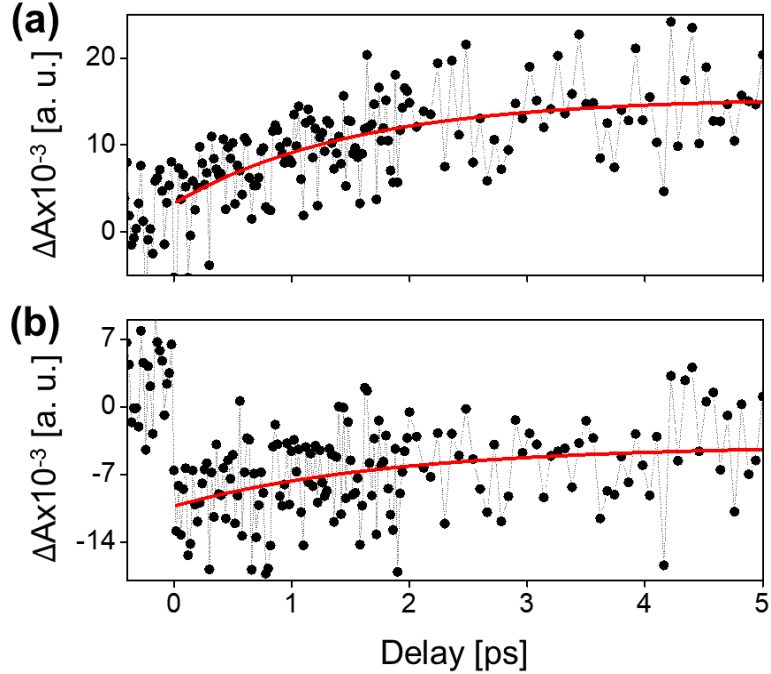

**Figure 8**. The temporal lineouts of averaged TA signals between (a) 54.5 and 55.1 eV, (b) 55.1 and 55.5 eV.

The correspondence of multiple timescales extracted here and at the W $O_3$ edge leads us to further compare the sources of those spectral features. In contrast to the W $O_3$ edge where both carrier state blocking and phonon excitations contribute to the XUV TA spectra, carriers do not contribute directly to the signals at the Se $M_5$ edge and phonon effects dominate. While the respective 1.6 ps and 1.7 ps rise here and at the W $O_3$ edge, respectively, can be both assigned to phonon population increase with possible contribution from non-radiative carrier recombination, the 0.4 ps dynamics arise from different degrees of freedom between the two spectral regions. At the W $O_3$ edge, the 0.4 ps shift reflects the holes relaxing to the valence band edge, whereas at the Se $M_5$ edge, the signal is not directly due to carriers and we attribute it to phonon-induced band renormalization from carrier-phonon scattering. The comparison of XUV TA spectra at the W $O_3$ edge and Se $M_5$ edge provides a look at the carrier-phonon scattering process from both electronic and phonon degrees of freedom, which cannot be easily obtained through other experimental methods such as electron diffraction.

Contrary to the positive feature at ~54.8 eV, the negative feature (55.1-55.5 eV) has a maximum amplitude at time zero and decays to a constant amplitude at long time delays. This suggests that this signal may be subjected to both carrier-induced band renormalization, which maximizes at zero time delay, and phonon effects, which become pronounced at the long time limit. The amplitude of this feature can be fitted with a single exponential decay of 2.2 ± 0.5 ps (Figure 8b). However, due to the low signal-to-noise ratio and the potential involvement of both carrier and phonon dynamics in the same spectral region, we do not attribute any new dynamics to this timescale.

Lastly, we compare carrier and phonon dynamics between $WSe_2$ and $WS_2$, which features highly similar XUV TA spectra.[19] Compared to $WS_2$, $WSe_2$ shows significantly faster hole relaxation. To understand the discrepancy, we first consider whether this difference in the hole relaxation time can be understood by the cooling process of high energy C band excitons through the band nesting effect,[45,46] as the energy gap between the C and B excitonic bands is much smaller in $WSe_2$ than $WS_2$.[45] However, the C exciton band edge is 2.4 eV in $WSe_2$ and 2.8 eV in $WS_2$. The latter is just on the edge or out of our broadband photoexcitation pulse ranging 1.23~2.45 eV. Thus, we can exclude the contribution from hot carrier relaxation arising from the C exciton.

We next consider whether the difference in the cooling time can be comprehended with their different phonon contributions to the hot carrier relaxation. For hot carrier cooling in a series of TMDs, Danovich *et al.* estimated phonon scattering rates with deformation potential interaction between carriers and phonons as a main energy dissipation path of hot carriers, and predicted that $WS_2$ exhibits a longer carrier cooling time than $WSe_2$.[47] In addition, it has been shown that the phonon scattering rates are inversely proportional to the phonon frequency and the reduced mass or overall atomic mass in the unit cell, depending on the phonon modes. Therefore, despite their similarities in crystal structures and various electronic properties, the faster hole-phonon relaxation in $WSe_2$ can be explained as it is more massive than $WS_2$ and features phonons with lower frequencies.

**Conclusion**

In summary, carrier and phonon dynamics in a multilayer $WSe_2$ film are investigated by XUV TA spectroscopy. The core-to-valence transition by XUV photoexcitation enables the extraction of hole dynamics directly from the observation of the W 5p core-to-VB transition, which illustrates hole recombination within a few picoseconds. The complementary observation of another part of the W and the Se edges reveals that a phonon-mediated band shift occurs in the same temporal range with the hole depopulation, which is well matched with the previously observed phonon population rise by femtosecond electron diffraction. This study provides a detailed first look on photoinduced dynamics with core-level excitations from both the metal and chalcogen atom in a W-based TMD, and this experimental approach can be further applied to investigate photophysical and photochemical processes in TMD multilayers and superlattices.

ASSOCIATED CONTENT

**Supporting Information**. Details for simulating transient absorption spectra, X-ray photoelectron spectra and Raman spectra, and coherent phonon motion at the selenium edge.

AUTHOR INFORMATION

**Corresponding Author**

Stephen R. Leone − Department of Chemistry and Department of Physics, University of California, Berkeley, California 94720, United States; Chemical Sciences Division, Lawrence Berkeley National Laboratory, Berkeley, California 94720, United States; orcid.org/0000-0003-1819-1338; Email: srl@ berkeley.edu

**Author Contributions**

[†]J. Oh and H.-T. Chang contributed equally to this work

**Notes**

Any additional relevant notes should be placed here.

ACKNOWLEDGMENT

Investigations were supported by the U.S. Air Force Office of Scientific Research Grants No. FA9550-19-1-0314, No. FA9550-20-1-0334, No. FA9550-15-0037 (concluded), and No. FA9550-14-1-0154 (concluded), the Defense Advanced Research Projects Agency PULSE Program Grant No. W31P4Q-13-1-0017 (concluded), the Army Research Office Grant No. W911NF-14-1-0383 (concluded), and W. M. Keck Foundation Award No. 046300-002. This research used resources of the Molecular Foundry (#6577) and the Advanced Light Source, U.S. DOE Office of Science User Facilities, under Contract No. DE-AC02-05CH11231. Core-level absorption simulations were conducted at Molecular Graphics and Computation Facility, UC Berkeley College of Chemistry, funded by National Institute of Health (NIH Grant No. S10OD023532), and the Scientific Compute Cluster at GWDG Göttingen, the joint data center of the Max Planck Society for the Advancement of Science (MPG) and University of Göttingen. H.-T.C. acknowledges support from Air Force Office of Scientific Research (AFOSR) (Grants No. FA9550-15-1-0037 and No. FA9550-19-1-0314) and the W. M. Keck Foundation (Grant No. 046300-002). J.O. is supported by the NRF funded by the Ministry of Science and ICT (2021R1C1C1013828 and 2021R1A6A1A03039503), and the Korean Basic Science Institute (National research Facilities and Equipment Center) grant (2022R1A6C101B794), with supplements from the AFOSR grants. This work was also supported by the Soonchunhyang University Research Fund.

**TOC Graphic**

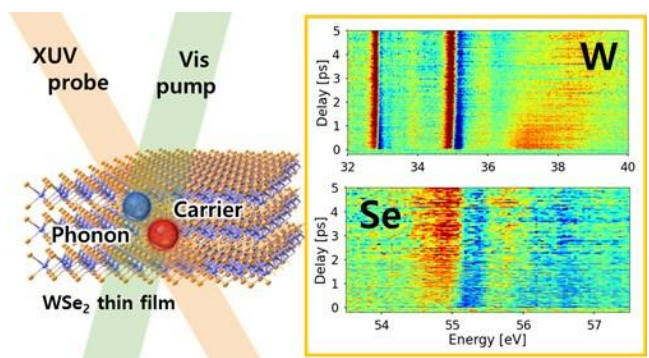